\documentclass[%
 superscriptaddress, amsmath, amssymb, aps, reprint, makecell, 
]{revtex4-2}

\usepackage{graphicx} 
\usepackage{dcolumn} 
\usepackage{bm} 
\usepackage{xcolor} 
\usepackage{hyperref}

\begin{document}

\preprint{APS/123-QED}

\title{Observation of Declination Dependence in the Cosmic Ray Energy Spectrum}

\author{R.U.~Abbasi}
\affiliation{Department of Physics, Loyola University Chicago, Chicago, Illinois 60660, USA}

\author{T.~Abu-Zayyad}
\affiliation{High Energy Astrophysics Institute and Department of Physics and Astronomy, University of Utah, Salt Lake City, Utah 84112-0830, USA}
\affiliation{Department of Physics, Loyola University Chicago, Chicago, Illinois 60660, USA}

\author{M.~Allen}
\affiliation{High Energy Astrophysics Institute and Department of Physics and Astronomy, University of Utah, Salt Lake City, Utah 84112-0830, USA}

\author{J.W.~Belz}
\affiliation{High Energy Astrophysics Institute and Department of Physics and Astronomy, University of Utah, Salt Lake City, Utah 84112-0830, USA}

\author{D.R.~Bergman}
\affiliation{High Energy Astrophysics Institute and Department of Physics and Astronomy, University of Utah, Salt Lake City, Utah 84112-0830, USA}

\author{I.~Buckland}
\affiliation{High Energy Astrophysics Institute and Department of Physics and Astronomy, University of Utah, Salt Lake City, Utah 84112-0830, USA}

\author{W.~Campbell}
\affiliation{High Energy Astrophysics Institute and Department of Physics and Astronomy, University of Utah, Salt Lake City, Utah 84112-0830, USA}

\author{B.G.~Cheon}
\affiliation{Department of Physics and The Research Institute of Natural Science, Hanyang University, Seongdong-gu, Seoul 426-791, Korea}

\author{K.~Endo}
\affiliation{Graduate School of Science, Osaka Metropolitan University, Sugimoto, Sumiyoshi, Osaka 558-8585, Japan}

\author{A.~Fedynitch}
\affiliation{Institute of Physics, Academia Sinica, Taipei City 115201, Taiwan}
\affiliation{Institute for Cosmic Ray Research, University of Tokyo, Kashiwa, Chiba 277-8582, Japan}

% ...

\collaboration{The Telescope Array Collaboration}
\noaffiliation

\date{\today}

\begin{abstract}
We report on an observation of the difference between northern and southern skies of the ultrahigh energy cosmic ray energy spectrum with a significance of ${\sim}8\sigma$.  We use measurements from the two largest experiments---the Telescope Array observing the northern hemisphere and the Pierre Auger Observatory viewing the southern hemisphere. Since the comparison of two measurements from different observatories introduces the issue of possible systematic differences between detectors and analyses, we validate the methodology of the comparison by examining the region of the sky where the apertures of the two observatories overlap. Although the spectra differ in this region, we find that there is only a $1.8\sigma$ difference between the spectrum measurements when anisotropic regions are removed and a fiducial cut in the aperture is applied.

\end{abstract}

\maketitle


\section{Introduction}
\label{sec:intro}
Ultrahigh energy cosmic rays (UHECRs) are believed to be charged particles with energies greater than $10^{18}$~eV, originating from outer space. Examining their energy spectrum is crucial because the features in the spectrum provide information on their potential sources and their propagation across the universe. An example of this is the high energy cutoff, first found by the High Resolution Fly's Eye experiment (HiRes) \cite{HiRes:2007lra} and later confirmed by Pierre Auger Observatory (Auger) \cite{PierreAuger:2008rol} and Telescope Array (TA) experiment \cite{TelescopeArray:2012qqu}. The TA \cite{TelescopeArray:2012uws, Tokuno:2012mi} and Auger \cite{PierreAuger:2004naf,PierreAuger:2015eyc} are currently the two largest UHECR observatories in operation. TA observes the northern hemisphere, while Auger views the southern hemisphere. Both observatories consist of fluorescence detectors (FDs) and surface detectors (SDs). Due to the fact that the FDs only operate on clear moonless nights, the SD data has the highest number of events by about an order of magnitude. For this reason, SD data is preferred for spectral and anisotropy studies.

In this work, we present the difference between the TA and Auger spectra at the highest energies, which has an ${\sim}8\sigma$ significance. This result is surprising and its validation is necessary. 
The TA spectrum views the declination region $-15.7^\circ < \delta < +90^\circ$, while the Auger spectrum observes $-90^\circ < \delta < +24.8^\circ$. Therefore, there is region of overlapping view between $-15.7^\circ$ and $+24.8^\circ$, which we call the {\it common declination band}. In this region, one would expect that the TA and Auger measurements should agree. However, this is true only when the energy spectra are independent of declination, and the apertures of the two experiments are identical. We will discuss the impact of these effects in this paper.

In the following sections, we provide an overview of the TA SD systems in Section~\ref{sec:TASD}, detail the datasets utilized for this study in Section~\ref{sec:data}, and present our results in Section~\ref{sec:results}. We describe the spectra in the common declination band and comment on anisotropy regions in Section~\ref{app:commonDec} and provide a summary in Section~\ref{sec:summary}. Finally, Appendix~\ref{app:datalist} presents the cosmic ray flux measured by the TA SD systems.

\section{Telescope Array Surface Detector}
\label{sec:TASD}

The Telescope Array is located near the city of Delta, Utah, USA, in the west desert at coordinates ($39.3^{\circ}$ N, $112.9^{\circ}$ W), with an elevation of 1400 m above sea level. The TA SD array  \cite{TelescopeArray:2012uws} consists of 507 scintillation detectors arranged in a square grid with a spacing of 1.2 km, covering an area of 700 km$^2$. Each detector consists of two layers of 1.2 cm thick plastic scintillator, stacked one above the other, and has an area of 3 m$^2$.  

When a cosmic ray air shower strikes the SD, there are thus two measurements of each detector’s pulse area.   Detectors are powered by solar cells and batteries, and radio towers communicate with the detectors.  The readout system consists of a flash analog-to-digital converter (FADC) with a 50 MHz sampling rate.  Calibration events consist of single muon hits, and their pulse area distributions are collected over 10-minute time intervals.  This allows every counter to be calibrated in terms of minimum-ionizing particles (MIPs) on a continual basis.  When three or more nearest neighbor detectors have pulse areas greater than 3 MIPs, within an 8 $\mu$s period, the array is triggered and each counter with signal greater than 0.3 MIPs reports its FADC waveforms to the communication tower. 

The reconstruction of cosmic ray properties is performed by two fitting procedures---time fit and lateral distribution fit. First, we utilize the modified Linsley shower-shape function \cite{Teshima:1986rq} to fit to the time distribution of the struck counters. This time fit yields the event's arrival direction and core position. Next, we perform a fit to the particle density distribution as a function of the distance from the shower axis, using the same lateral distribution function as employed by the AGASA experiment \cite{Yoshida:1994jf,Takeda:2002at}. From this lateral distribution fit, we interpolate the density of shower particles at a lateral distance of 800 m from the shower axis, denoted as S(800).

Using S(800) and the zenith angle of the incident cosmic-ray arrival direction, the cosmic ray’s energy is determined from a look-up table calculated using a Monte Carlo (MC) simulation of the experiment \cite{TelescopeArray:2014nxa,Stokes:2011wf} \footnote{The MC described here uses events generated by the CORSIKA simulation package \cite{Heck:1998vt} using the QGSJET-II-03 high-energy hadronic interaction model \cite{Ostapchenko:2004ss} with an assumption of proton primaries. Since, in the end, we normalize the energy scale to that of the FD, the spectrum we calculate here is insensitive to the assumption of primary particles or the use of various available hadronic interaction models.}.  This SD energy determination may have potential biases linked to the modeling of hadronic interactions in MC simulations. In contrast, an FD's energy measurement is calorimetric, and as a result, their energy scale uncertainty is experimentally well controlled. Therefore, we normalize the SD energy scale to that of the FD by utilizing events observed by both detectors. It was determined that the energy scale of the SD is 27\% higher than that of the FD, independent of energy \cite{TelescopeArray:2012qqu}.  Therefore, a 27\% normalization in SD energy determined by the MC simulation is performed. 

In addition, the constant intensity cut (CIC) method has also been used to determine the cosmic ray energy.  This analysis was designed to be almost identical to that of Auger, taking into account attenuation of the shower in the atmosphere \cite{PierreAuger:2020qqz}.  The CIC energy scale is again normalized by FD measurements.  We compared the energies determined by the TA standard method, using the MC look-up table including the energy scaling to FD energy, with those obtained through the CIC method (which is independent of MC). It is found that the CIC energies agree within 2\% with those determined by the TA standard method \cite{Kim:2023eul,TelescopeArray:2023bdy}.

For comparison, Auger is located near the town of Malargüe, Mendoza, Argentina, at coordinates ($35.2^{\circ}$ S, $69.4^{\circ}$  W), with an elevation of 1400 m above sea level \cite{PierreAuger:2004naf,PierreAuger:2015eyc}. The Auger SD consists of large water Cherenkov detectors, placed in a triangular grid of 1.5 km spacing with an area of about 3000 km$^2$.  The spectrum is calculated using only the energy range where the detector is nearly 100\% efficient, and a MC simulation is only used to correct for bin-to-bin migration of events (which is largest at the highest energies).

\section{Datasets}
\label{sec:data}

For this work, we utilized TA data collected between May 11, 2008, and May 10, 2022. For comparison, we employed Auger ``vertical" events (zenith angle less than 60$^\circ$) as shown in \cite{PierreAuger:2020qqz}. To the TA data, we applied event selection criteria as explained below: 
\begin{enumerate}
  \setlength{\itemsep}{0pt}
  \setlength{\parskip}{0pt}
    \item Each event must include at least five SD counters.
    \item The reconstructed zenith angle must be less than 55$^\circ$.
    \item Both the geometry and lateral distribution fits must have $\chi^2$/degree of freedom less than 4.
    \item The angular uncertainty estimated by the geometry fit must be less than 5$^\circ$.
    \item The fractional uncertainty in S(800) estimated by the lateral distribution fit must be less than 25\%.
    \item The counter with the largest signal must be surrounded by four working counters: one to the north, east, south, and west on the grid, but they do not have to be immediate neighbors of the largest signal counter.
\end{enumerate}

In our previous paper on the energy spectrum measurements \cite{TelescopeArray:2012qqu}, we applied event selection criteria with slightly different cuts aimed at optimizing energy resolution. However, the selection criteria described above employs a slightly looser set of cuts than in \cite{TelescopeArray:2012qqu} in order to maximize data statistics in high energy regions. Notable differences include zenith angles less than 55$^\circ$ and energies greater than $10^{18.8}$~eV, where the detector is almost 100\% efficient \cite{TelescopeArray:2012qqu}. These criteria were initially selected to increase data statistics for anisotropy studies while keeping reasonable energy and angular resolutions, but we later adopted them for the TA and Auger Joint Spectrum Working Group's studies to maximize statistics in high energy regions as well as in the declination region seen by both experiments. With these selection criteria, we have 12,845 events with energies greater than $10^{18.8}$~eV in the dataset.

\section{Results}
\label{sec:results}

\begin{figure}[btp]
\includegraphics[width=1\columnwidth]{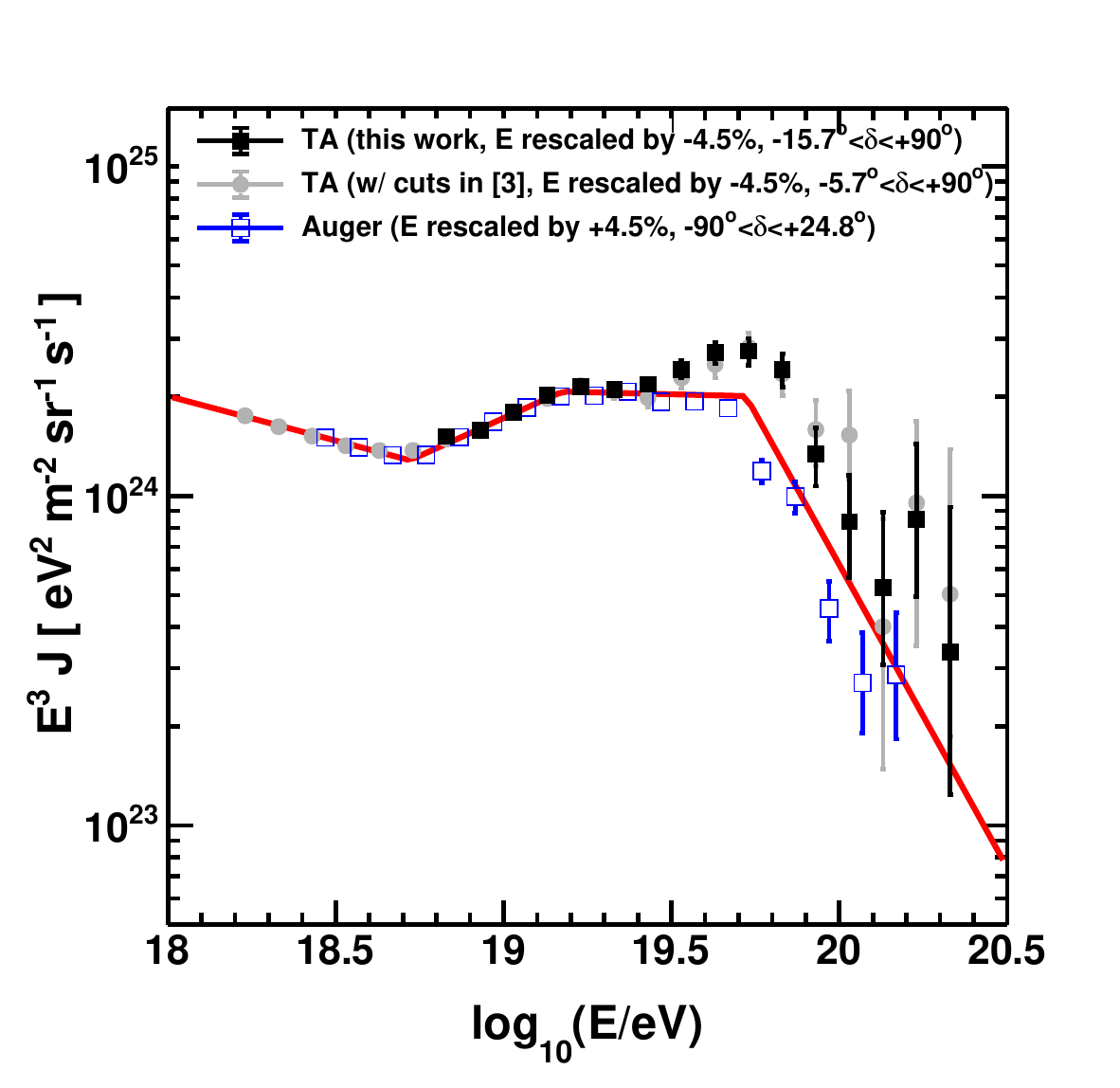}
\caption{\label{fig:joinFit_fullsky} Energy spectra of TA and Auger measured in their whole apertures. The black full squares indicate the energy spectrum of TA within the declination range of $-15.7^{\circ}$ to $+90^{\circ}$, which used the cuts described in Section~\ref{sec:data}. The blue open squares represent the energy spectrum of Auger spanning $-90^{\circ}$ to $+24.8^{\circ}$ in declination. For reference, TA data with the cuts described in~\cite{TelescopeArray:2012qqu} is shown using the gray full circles. Its full aperture is within the declination range of $-5.7^{\circ}$ to $+90^{\circ}$. The energy scale difference has been corrected by raising the Auger energy scale by 4.5\% and lowering the TA energy scale by 4.5\%. After adjusting the overall energy scale, the two spectra agree well up to $10^{19.5}$ eV. The red line represents the result of the joint fit spectra between TA (black full squares) and Auger (blue open squares). 
}
\end{figure}

Figure~\ref{fig:joinFit_fullsky} shows the spectra of TA and Auger, adjusted for the overall energy scale by raising Auger's energy scale by 4.5\% and lowering TA's by 4.5\%. This 9\% overall energy scale difference between the two measurements is well understood thanks to the efforts of the TA and Auger Joint Spectrum Working Group, which was established to investigate differences in spectrum measurements. It arises from the use of different constants in the reconstruction of fluorescence data and these different constants yield a negligible energy dependence \cite{Verzi:2017hro, Deligny:2020gzq, TelescopeArray:2021zox, Tsunesada:2023yhw}.

In Figure~\ref{fig:joinFit_fullsky}, the black full squares indicate the energy spectrum of TA within the declination range of $-15.7^{\circ}$ to $+90^{\circ}$, which used the cuts described in Section~\ref{sec:data}. The blue open squares represent the energy spectrum of Auger spanning $-90^{\circ}$ to $+24.8^{\circ}$ in declination. Note that the gray full circles indicate the TA data selected based on the criteria outlined in \cite{TelescopeArray:2012qqu} to encompass as wide an energy range as possible. For energies below $10^{18.8}$~eV the TA SD does not have 100\% efficiency, and a correction has been made by Monte Carlo calculation. The comparison shows that the spectrum measurements by TA and Auger align for energies below about $10^{19.5}$~eV, above which a growing disagreement becomes evident. The high-energy cutoff occurs at different energies in the two hemispheres. 

To quantify the level of agreement or disagreement between the two spectra, we performed a joint fit to both cosmic ray spectra into a broken power law function (power law segments with three break points) using the binned Poisson likelihood method, Eq. 39.16 in \cite{Patrignani_2016}. This fit takes into account the numbers of events, the exposure, and the resolution correction factors of both experiments. The red line in Figure~\ref{fig:joinFit_fullsky} represents the result of this joint fit for data from TA (shown as black full squares) and Auger (shown as blue open squares). The cosmic ray flux measured by the TA SD for this study, utilized in Figure~\ref{fig:joinFit_fullsky}, is provided in Appendix~\ref{app:datalist}. From the log-likelihood sum over event bins for the joint fit, we calculate the significance of the spectrum difference. The fit gave the log-likelihood sum of 130.33 for 26 degrees of freedom, corresponding to a Poisson probability of $7.5\times10^{-16}$. This corresponds to a one-sided test significance of $8.0\sigma$.

\section{Spectra in the Common Declination Band}
\label{app:commonDec}

\begin{figure}[t]
\includegraphics[width=1\columnwidth]{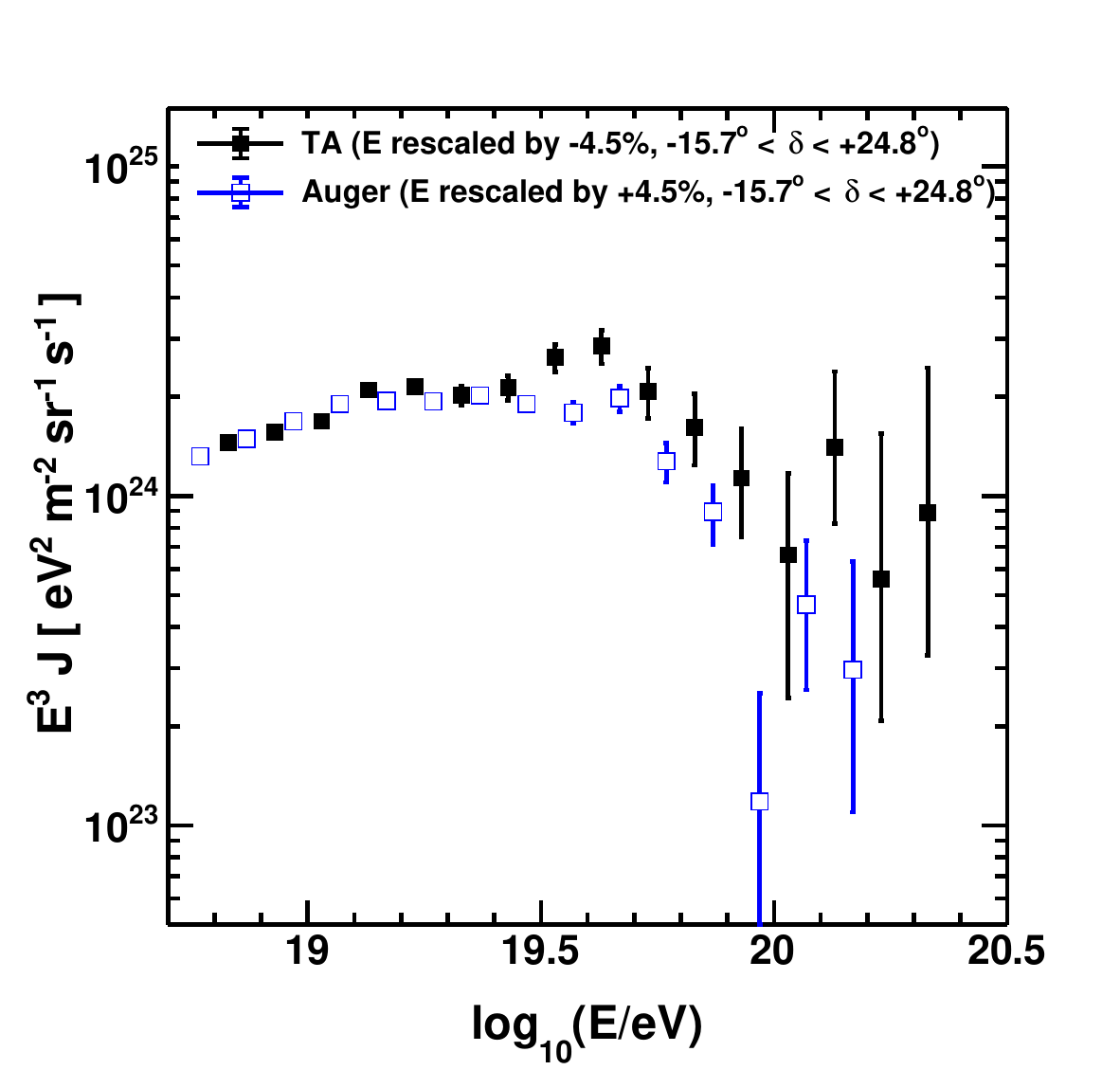}
\caption{\label{fig:spec_common} TA (black full squares) and Auger (blue open squares) spectra in the full common declination band $-15.7^{\circ} < \delta < +24.8^{\circ}$. 
}
\end{figure}

\begin{figure}
\includegraphics[width=0.95\columnwidth]{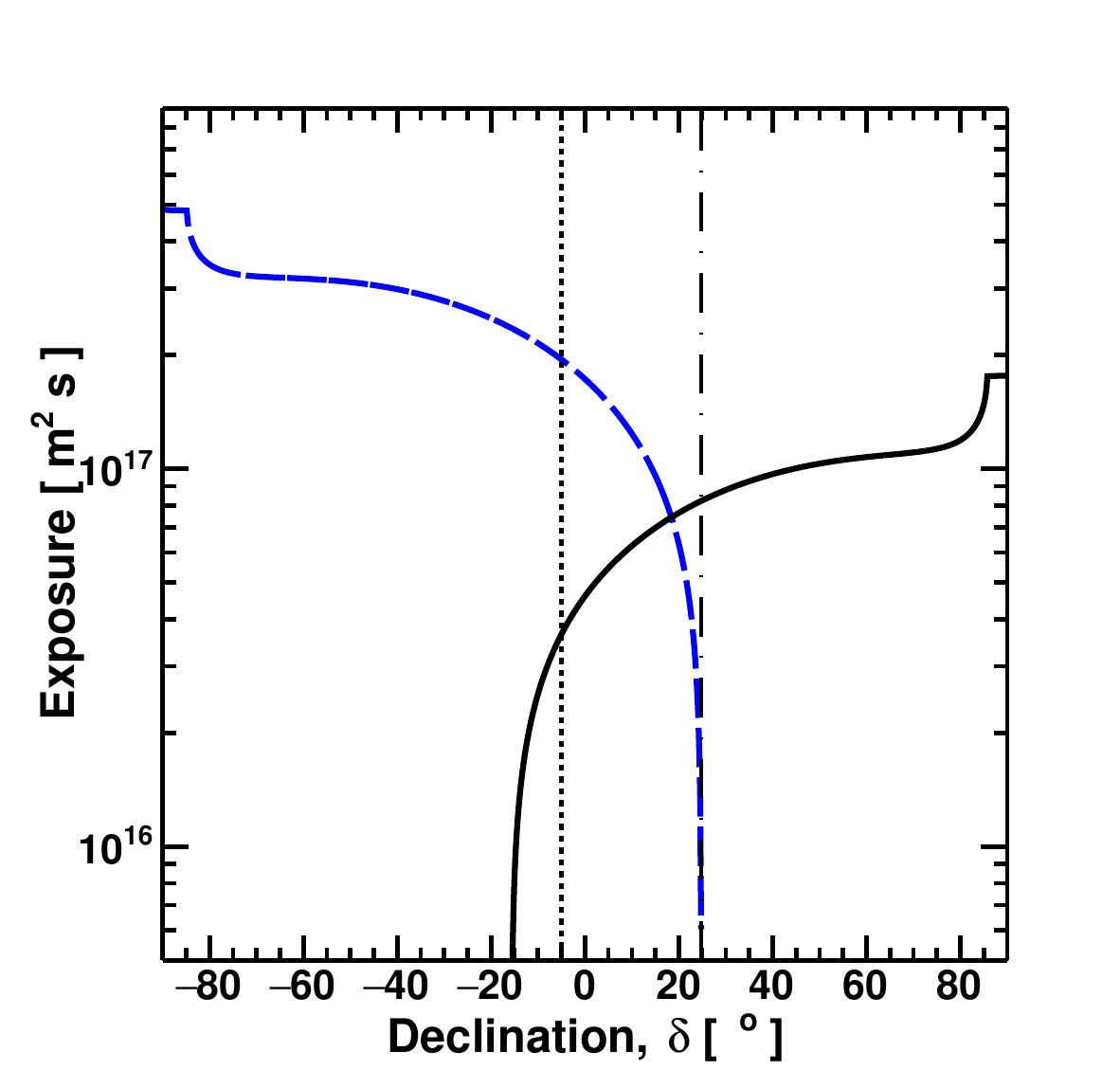}
\caption{\label{fig:exposure} TA and Auger exposures as a function of declination.  The black solid line represents the TA exposure, and the blue dashed line indicates the Auger exposure. The dotted vertical line corresponds to a declination of -5$^\circ$, while the dash-dotted vertical line indicates a declination of +24.8$^\circ$.
}
\end{figure}

\begin{figure}
\includegraphics[width=1\columnwidth]{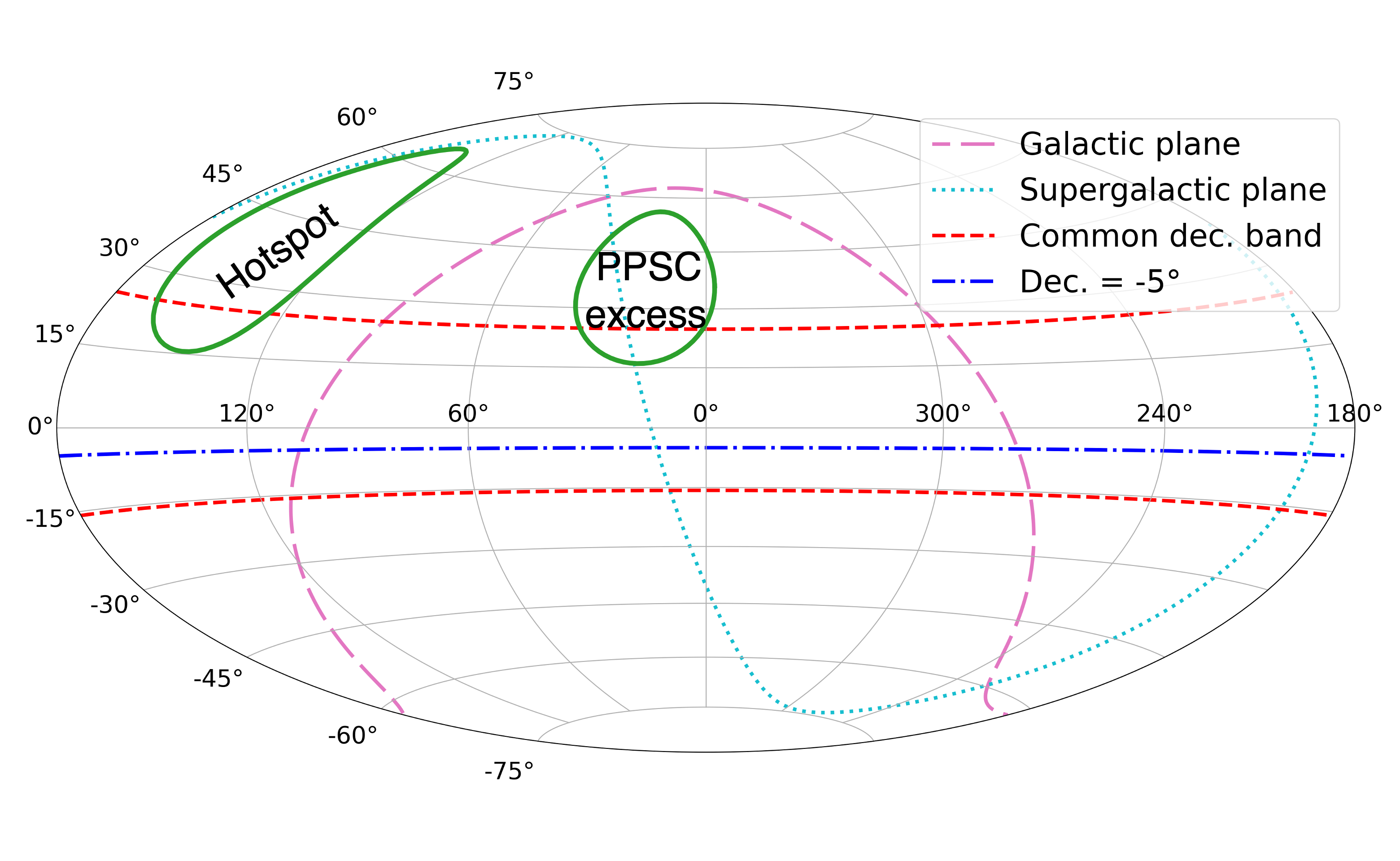}
\caption{\label{fig:skymaps} Sky map in equatorial coordinates using Hammer projection. The green circles indicate the locations of the Hotspot and the Perseus-Pisces supercluster (PPSC) excess regions, respectively. The two red dashed lines indicate the edges of the common declination band.  The blue dash-dotted line represents the fiducial cut location at a declination of -5$^\circ$.  
}
\end{figure}

\begin{figure}
\includegraphics[width=1\columnwidth]{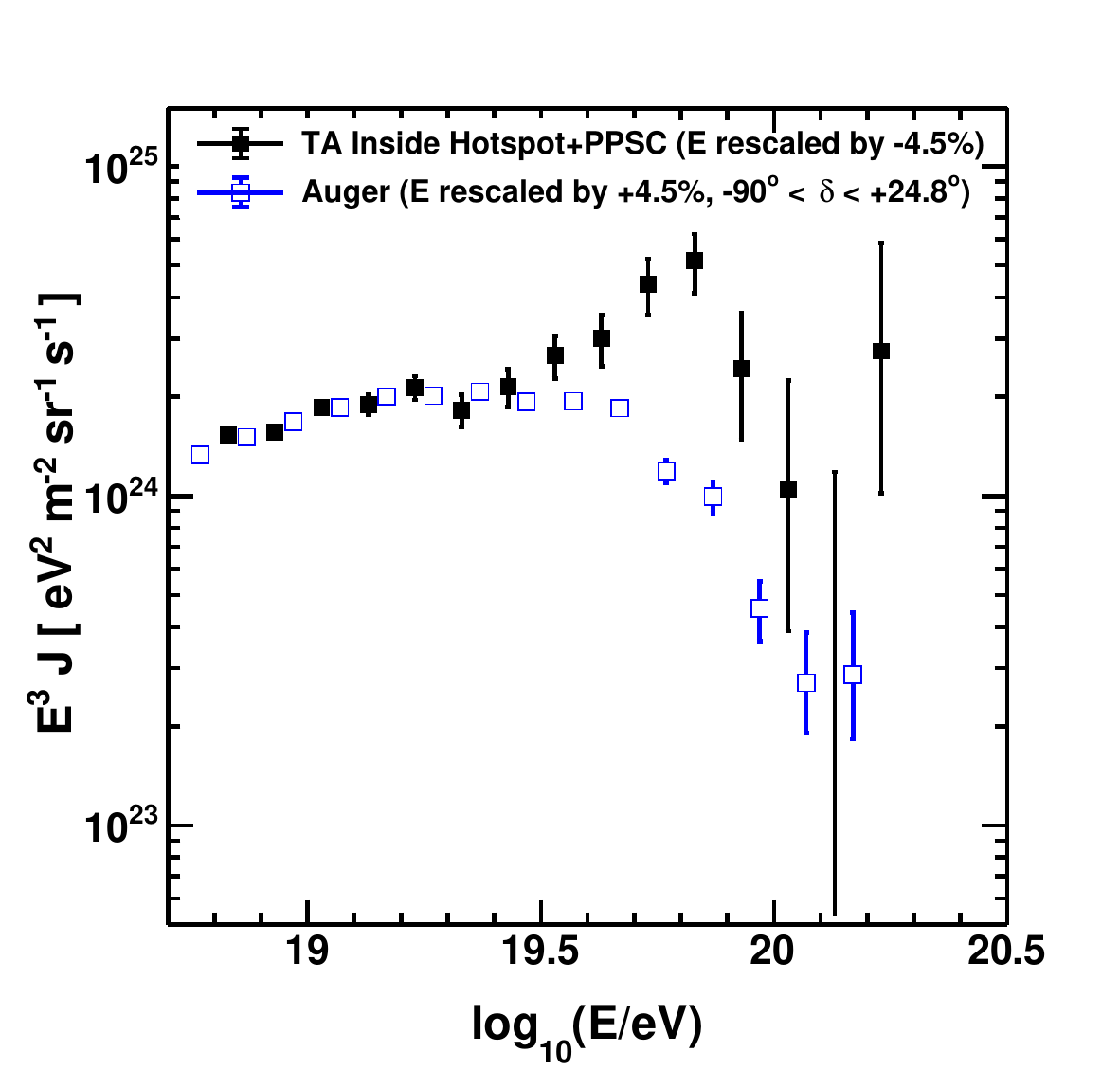}
\caption{\label{fig:spec_inside} The black full squares indicate the spectrum of events inside the Hotspot and PPSC excess regions, depicted with the green circles in Figure~\ref{fig:skymaps}. For reference, the Auger spectrum in their full aperture is displayed with the blue open squares.
}
\end{figure}

\begin{figure}
\includegraphics[width=1\columnwidth]{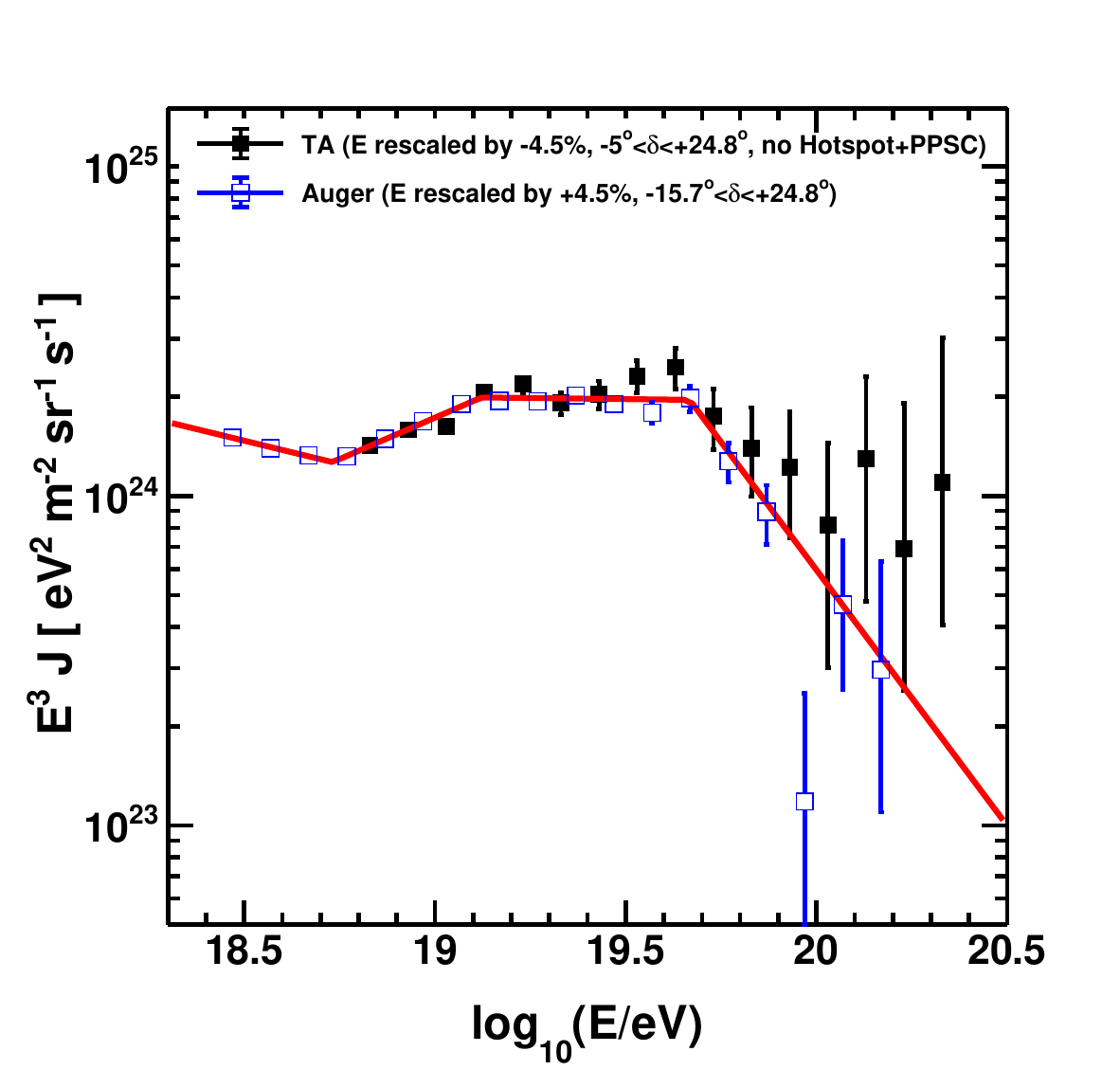}
\caption{\label{fig:common_restricted} Joint fit spectra comparison between TA and Auger in the common declination band with the described cuts applied to the TA data. The black full squares indicate the TA data after the fiducial cut in the aperture is applied and anisotropic regions are removed, while the blue open squares represent the Auger data. The red line depicts the result of the joint fit.
}
\end{figure}

TA and Auger have different types of surface detectors, use somewhat different reconstruction techniques, and their apertures have different declination dependence in the common declination band. Comparing their spectra in this region of the sky is a stringent test of whether they have comparable results. Figure~\ref{fig:spec_common} shows the spectra of the two experiments in the declination band $-15.7^\circ < \delta < +24.8^\circ$. The Auger data within the common declination band was utilized, as shown in \cite{Tsunesada:2023yhw}. The spectra seem to disagree at energies greater than $10^{19.5}$~eV. 

To understand this discrepancy, we revisited the analysis by introducing the most direct comparison possible of the spectra from TA and Auger within this band. First, we chose to implement a fiducial declination cut in the TA data. Figure~\ref{fig:exposure} shows the TA and Auger exposures as a function of declination \cite{Sommers:2000us}. The black solid line represents TA exposure, and the blue dashed line indicates Auger exposure. Notably, the exposure of TA at its southernmost edge changes extremely rapidly. Therefore, we implemented the fiducial cut requiring $\delta > -5^\circ$ (the black dotted vertical line in Figure~\ref{fig:exposure}) to avoid this region of the sky. This cut excluded 654 events out of a total of 4,861 events in the common declination band.

Another notable point is the difference between the sky just north of the common declination band and that to the south, as TA data shows anisotropy regions. These include anisotropy signals such as the Hotspot and the Perseus-Pisces supercluster (PPSC) excess, which were identified through oversampling searches using intermediate-scale angular circles \cite{TelescopeArray:2014tsd,TelescopeArray:2021dfb,Kim:2023ksw}. Figure~\ref{fig:skymaps} shows a sky map in equatorial coordinates using the Hammer projection to depict the locations of these excess regions. The two red dashed horizontal lines are the boundaries of the common declination band at $\delta = -15.7^\circ$ and $+24.8^\circ$. Additionally, we mark the location of the fiducial cut at $\delta = -5^\circ$ (the area below the blue line is cut out) with the blue dash-dotted horizontal line. The two green circles indicate the Hotspot and PPSC excess regions in the TA data~\footnote{
Note that the anisotropy signal regions depicted are based on the previous analysis results as follows. The Hotspot was identified in events with energies exceeding $\sim$$10^{19.75}$~eV, located at equatorial coordinates ($144.0^{\circ}, 40.5^{\circ}$) within a $25^{\circ}$ radius. Additionally, we observed additional anisotropies in events with energies greater than $10^{19.4}$~eV in the direction of the Perseus-Pisces supercluster. The PPSC excess was located at ($17.9^{\circ}, 35.2^{\circ}$) in equatorial coordinates within a $20^{\circ}$ radius.
}. 

Both excess regions extend down into the common declination band. However, Auger has not reported any anisotropy regions intruding into the common declination band from the south \cite{PierreAuger:2022axr}. Notably, the two TA excess regions in the common declination band are close to the northernmost edge of Auger's exposure, where it is rapidly falling. (See the blue dashed line in Figure~\ref{fig:exposure}.) We adopt the hypothesis that the TA excesses may affect the spectral characteristics observed within the common declination band. This influence could be significant if the spectrum within the anisotropy regions differs from that of the background. Figure~\ref{fig:spec_inside} shows the spectrum of events inside the Hotspot and PPSC excess regions, supporting that this is indeed the case. Therefore, we excluded 269 events from these excess regions out of a total of 4,861 events in the common declination band and reanalyzed the spectrum. We aimed to make the most direct comparison of the spectra from the TA and Auger within this band.

Figure~\ref{fig:common_restricted} displays the results of a joint fit to the TA and Auger spectra, depicted by the red line, using data from the common declination band and after applying the cuts described above to the TA data. The black full squares indicate the TA data from the common declination band, following the fiducial cut in the aperture at $\delta > -5^\circ$ and the removal of the two anisotropic regions, while the Auger data within the common declination band are represented by the blue open squares. The fit yielded the log-likelihood sum of 40.12 for 26 degrees of freedom, corresponding to a Poisson probability of $3.8\times10^{-2}$. This is equivalent to a one-sided test significance of $1.8\sigma$.  Therefore, there is no statistically significant difference between the spectra. This constitutes a validation of the analysis methods of TA and Auger.  Once comparable data sets are selected, the results are consistent within statistics.

\section{Summary}
\label{sec:summary}
The spectrum difference between TA and Auger has long been a source of controversy. How could two experiments have spectra that agree very well below $10^{19.5}$~eV, then disagree so much above this energy?  TA sees a more intense flux of cosmic rays and a higher cutoff energy.  The two collaborations have founded a Spectrum Working Group to investigate differences, which clarified the origin of the overall energy scale difference to be in the fluorescence yield and other constants used in setting the energy scales of both experiments. Under the Working Group auspice, a study of the common declination band was initiated.  After the analysis described in Section~\ref{app:commonDec}, we find that the TA and Auger spectra in the common declination band are in agreement within $1.8\sigma$.
 
Having validated the TA and Auger spectrum calculation methods, we quantify the declination dependence of the spectra as seen in the whole apertures of TA and Auger.  A joint fit to the two spectra was performed, and the log-likelihood per degree of freedom was found to be $8.0\sigma$.  This constitutes the observation that the UHECR spectrum differs in the northern and southern hemispheres.  We show that a significant part of the difference is due to events from the Hotspot and Perseus-Pisces supercluster excess regions.

\section*{Acknowledgements}

The Telescope Array experiment is supported by the Japan Society for
the Promotion of Science(JSPS) through
Grants-in-Aid
for Priority Area
431,
for Specially Promoted Research
JP21000002,
for Scientific  Research (S)
JP19104006,
for Specially Promoted Research
JP15H05693,
for Scientific  Research (S)
JP19H05607,
for Scientific  Research (S)
JP15H05741,
for Science Research (A)
JP18H03705,
for Young Scientists (A)
JPH26707011,
and for Fostering Joint International Research (B)
JP19KK0074,
by the joint research program of the Institute for Cosmic Ray Research (ICRR), The University of Tokyo;
by the Pioneering Program of RIKEN for the Evolution of Matter in the Universe (r-EMU);
by the U.S. National Science Foundation awards
PHY-1806797, PHY-2012934, PHY-2112904, PHY-2209583, PHY-2209584, and PHY-2310163, as well as AGS-1613260, AGS-1844306, and AGS-2112709;
by the National Research Foundation of Korea
(2017K1A4A3015188, 2020R1A2C1008230, and 2020R1A2C2102800) ;
by the Ministry of Science and Higher Education of the Russian Federation under the contract 075-15-2024-541, IISN project No. 4.4501.18, by the Belgian Science Policy under IUAP VII/37 (ULB), by National Science Centre in Poland grant 2020/37/B/ST9/01821, by the European Union and Czech Ministry of Education, Youth and Sports through the FORTE project No. CZ.02.01.01/00/22\_008/0004632, and by the Simons Foundation (00001470, NG). This work was partially supported by the grants of the joint research program of the Institute for Space-Earth Environmental Research, Nagoya University and Inter-University Research Program of the Institute for Cosmic Ray Research of University of Tokyo. The foundations of Dr. Ezekiel R. and Edna Wattis Dumke, Willard L. Eccles, and George S. and Dolores Dor\'e Eccles all helped with generous donations. The State of Utah supported the project through its Economic Development Board, and the University of Utah through the Office of the Vice President for Research. The experimental site became available through the cooperation of the Utah School and Institutional Trust Lands Administration (SITLA), U.S. Bureau of Land Management (BLM), and the U.S. Air Force. We appreciate the assistance of the State of Utah and Fillmore offices of the BLM in crafting the Plan of Development for the site.  We thank Patrick A.~Shea who assisted the collaboration with much valuable advice and provided support for the collaboration’s efforts. The people and the officials of Millard County, Utah have been a source of steadfast and warm support for our work which we greatly appreciate. We are indebted to the Millard County Road Department for their efforts to maintain and clear the roads which get us to our sites. We gratefully acknowledge the contribution from the technical staffs of our home institutions. An allocation of computing resources from the Center for High Performance Computing at the University of Utah as well as the Academia Sinica Grid Computing Center (ASGC) is gratefully acknowledged.

\appendix

\section{Spectrum Data Points}
\label{app:datalist}
Table~\ref{tab:datalist} provides the cosmic ray flux for each energy bin depicted in Figure~\ref{fig:joinFit_fullsky}, utilizing 14 years of Telescope Array surface detector data, collected between May 11, 2008, and May 10, 2022, in the full aperture of $-15.7^{\circ} < \delta < +90^{\circ}$. Note that the energy values in Figure~\ref{fig:joinFit_fullsky} have been reduced by 4.5\% compared to those detailed here. Table~\ref{tab:datalist} includes $\log_{10} (E/\textrm{eV})$ representing the energy of the bin center, J denoting the flux in the unit of [eV$^{-1}$m$^{-2}$sr$^{-1}$s$^{-1}$], and $\sigma_{\rm upper}$ and $\sigma_{\rm lower}$ representing the statistical uncertainties on the flux, corresponding to the upper and lower 68\% confidence limits. All uncertainties are expressed in the unit of [eV$^{-1}$m$^{-2}$sr$^{-1}$s$^{-1}$].

\begin{table}[]
\caption{\label{tab:datalist} Spectrum data points. For each energy bin, J denotes the flux, and $\sigma_{\rm upper}$ and $\sigma_{\rm lower}$ represent the statistical uncertainties on the flux, corresponding to the upper and lower 68\% confidence limits. Their units are in [eV$^{-1}$m$^{-2}$sr$^{-1}$s$^{-1}$].}
\begin{ruledtabular}
\begin{tabular}{cccc}
$\log_{10} (E/\textrm{eV})$ & \textrm{J} & $\sigma_{\rm upper}$ & $\sigma_{\rm lower}$
\\
\colrule
18.85&	4.58$\times 10^{-33}$&	8.88$\times 10^{-35}$&	8.88$\times 10^{-35}$\\
18.95&	2.43$\times 10^{-33}$&	5.44$\times 10^{-35}$&	5.44$\times 10^{-35}$\\
19.05&	1.39$\times 10^{-33}$&	3.72$\times 10^{-35}$&	3.72$\times 10^{-35}$\\
19.15&	7.66$\times 10^{-34}$&	2.53$\times 10^{-35}$&	2.53$\times 10^{-35}$\\
19.25&	4.22$\times 10^{-34}$&	1.64$\times 10^{-35}$&	1.64$\times 10^{-35}$\\
19.35&	2.03$\times 10^{-34}$&	9.89$\times 10^{-36}$&	9.89$\times 10^{-36}$\\
19.45&	9.69$\times 10^{-35}$&	6.10$\times 10^{-36}$&	6.10$\times 10^{-36}$\\
19.55&	5.59$\times 10^{-35}$&	4.12$\times 10^{-36}$&	4.12$\times 10^{-36}$\\
19.65&	3.07$\times 10^{-35}$&	2.80$\times 10^{-36}$&	2.80$\times 10^{-36}$\\
19.75&	1.73$\times 10^{-35}$&	1.89$\times 10^{-36}$&	1.89$\times 10^{-36}$\\
19.85&	7.23$\times 10^{-36}$&	1.01$\times 10^{-36}$&	1.01$\times 10^{-36}$\\
19.95&	2.46$\times 10^{-36}$&	5.13$\times 10^{-37}$&	6.53$\times 10^{-37}$\\
20.05&	1.17$\times 10^{-36}$&	3.10$\times 10^{-37}$&	4.20$\times 10^{-37}$\\
20.15&	1.55$\times 10^{-37}$&	9.79$\times 10^{-38}$&	1.75$\times 10^{-37}$\\
20.25&	1.85$\times 10^{-37}$&	1.17$\times 10^{-37}$&	1.42$\times 10^{-37}$\\
20.35&	4.90$\times 10^{-38}$&	3.09$\times 10^{-38}$&	8.60$\times 10^{-38}$\\
\end{tabular}
\end{ruledtabular}
\end{table}


\begin{thebibliography}{29}%
\makeatletter
\providecommand \@ifxundefined [1]{%
 \@ifx{#1\undefined}
}%
\providecommand \@ifnum [1]{%
 \ifnum #1\expandafter \@firstoftwo
 \else \expandafter \@secondoftwo
 \fi
}%
\providecommand \@ifx [1]{%
 \ifx #1\expandafter \@firstoftwo
 \else \expandafter \@secondoftwo
 \fi
}%
\providecommand \natexlab [1]{#1}%
\providecommand \enquote  [1]{``#1''}%
\providecommand \bibnamefont  [1]{#1}%
\providecommand \bibfnamefont [1]{#1}%
\providecommand \citenamefont [1]{#1}%
\providecommand \href@noop [0]{\@secondoftwo}%
\providecommand \href [0]{\begingroup \@sanitize@url \@href}%
\providecommand \@href[1]{\@@startlink{#1}\@@href}%
\providecommand \@@href[1]{\endgroup#1\@@endlink}%
\providecommand \@sanitize@url [0]{\catcode `\\12\catcode `\$12\catcode
  `\&12\catcode `\#12\catcode `\^12\catcode `\_12\catcode `\%12\relax}%
\providecommand \@@startlink[1]{}%
\providecommand \@@endlink[0]{}%
\providecommand \url  [0]{\begingroup\@sanitize@url \@url }%
\providecommand \@url [1]{\endgroup\@href {#1}{\urlprefix }}%
\providecommand \urlprefix  [0]{URL }%
\providecommand \Eprint [0]{\href }%
\providecommand \doibase [0]{https://doi.org/}%
\providecommand \selectlanguage [0]{\@gobble}%
\providecommand \bibinfo  [0]{\@secondoftwo}%
\providecommand \bibfield  [0]{\@secondoftwo}%
\providecommand \translation [1]{[#1]}%
\providecommand \BibitemOpen [0]{}%
\providecommand \bibitemStop [0]{}%
\providecommand \bibitemNoStop [0]{.\EOS\space}%
\providecommand \EOS [0]{\spacefactor3000\relax}%
\providecommand \BibitemShut  [1]{\csname bibitem#1\endcsname}%
\let\auto@bib@innerbib\@empty
\bibitem [{\citenamefont {Abbasi}\ \emph {et~al.}(2008)\citenamefont {Abbasi}
  \emph {et~al.}}]{HiRes:2007lra}%
  \BibitemOpen
  \bibfield  {author} {\bibinfo {author} {\bibfnamefont {R.~U.}\ \bibnamefont
  {Abbasi}} \emph {et~al.} (\bibinfo {collaboration} {HiRes}),\ }\bibfield
  {title} {\bibinfo {title} {{First observation of the Greisen-Zatsepin-Kuzmin
  suppression}},\ }\href {https://doi.org/10.1103/PhysRevLett.100.101101}
  {\bibfield  {journal} {\bibinfo  {journal} {Phys. Rev. Lett.}\ }\textbf
  {\bibinfo {volume} {100}},\ \bibinfo {pages} {101101} (\bibinfo {year}
  {2008})},\ \Eprint {https://arxiv.org/abs/astro-ph/0703099}
  {arXiv:astro-ph/0703099} \BibitemShut {NoStop}%
\bibitem [{\citenamefont {Abraham}\ \emph {et~al.}(2008)\citenamefont {Abraham}
  \emph {et~al.}}]{PierreAuger:2008rol}%
  \BibitemOpen
  \bibfield  {author} {\bibinfo {author} {\bibfnamefont {J.}~\bibnamefont
  {Abraham}} \emph {et~al.} (\bibinfo {collaboration} {Pierre Auger}),\
  }\bibfield  {title} {\bibinfo {title} {{Observation of the suppression of the
  flux of cosmic rays above $4\times 10^{19}$eV}},\ }\href
  {https://doi.org/10.1103/PhysRevLett.101.061101} {\bibfield  {journal}
  {\bibinfo  {journal} {Phys. Rev. Lett.}\ }\textbf {\bibinfo {volume} {101}},\
  \bibinfo {pages} {061101} (\bibinfo {year} {2008})},\ \Eprint
  {https://arxiv.org/abs/0806.4302} {arXiv:0806.4302 [astro-ph]} \BibitemShut
  {NoStop}%
\bibitem [{\citenamefont {Abu-Zayyad}\ \emph
  {et~al.}(2013{\natexlab{a}})\citenamefont {Abu-Zayyad} \emph
  {et~al.}}]{TelescopeArray:2012qqu}%
  \BibitemOpen
  \bibfield  {author} {\bibinfo {author} {\bibfnamefont {T.}~\bibnamefont
  {Abu-Zayyad}} \emph {et~al.} (\bibinfo {collaboration} {Telescope Array}),\
  }\bibfield  {title} {\bibinfo {title} {{The Cosmic Ray Energy Spectrum
  Observed with the Surface Detector of the Telescope Array Experiment}},\
  }\href {https://doi.org/10.1088/2041-8205/768/1/L1} {\bibfield  {journal}
  {\bibinfo  {journal} {Astrophys. J. Lett.}\ }\textbf {\bibinfo {volume}
  {768}},\ \bibinfo {pages} {L1} (\bibinfo {year} {2013}{\natexlab{a}})},\
  \Eprint {https://arxiv.org/abs/1205.5067} {arXiv:1205.5067 [astro-ph.HE]}
  \BibitemShut {NoStop}%
\bibitem [{\citenamefont {Abu-Zayyad}\ \emph
  {et~al.}(2013{\natexlab{b}})\citenamefont {Abu-Zayyad} \emph
  {et~al.}}]{TelescopeArray:2012uws}%
  \BibitemOpen
  \bibfield  {author} {\bibinfo {author} {\bibfnamefont {T.}~\bibnamefont
  {Abu-Zayyad}} \emph {et~al.} (\bibinfo {collaboration} {Telescope Array}),\
  }\bibfield  {title} {\bibinfo {title} {{The surface detector array of the
  Telescope Array experiment}},\ }\href
  {https://doi.org/10.1016/j.nima.2012.05.079} {\bibfield  {journal} {\bibinfo
  {journal} {Nucl. Instrum. Meth. A}\ }\textbf {\bibinfo {volume} {689}},\
  \bibinfo {pages} {87} (\bibinfo {year} {2013}{\natexlab{b}})},\ \Eprint
  {https://arxiv.org/abs/1201.4964} {arXiv:1201.4964 [astro-ph.IM]}
  \BibitemShut {NoStop}%
\bibitem [{\citenamefont {Tokuno}\ \emph {et~al.}(2012)\citenamefont {Tokuno}
  \emph {et~al.}}]{Tokuno:2012mi}%
  \BibitemOpen
  \bibfield  {author} {\bibinfo {author} {\bibfnamefont {H.}~\bibnamefont
  {Tokuno}} \emph {et~al.},\ }\bibfield  {title} {\bibinfo {title} {{New air
  fluorescence detectors employed in the Telescope Array experiment}},\ }\href
  {https://doi.org/10.1016/j.nima.2012.02.044} {\bibfield  {journal} {\bibinfo
  {journal} {Nucl. Instrum. Meth. A}\ }\textbf {\bibinfo {volume} {676}},\
  \bibinfo {pages} {54} (\bibinfo {year} {2012})},\ \Eprint
  {https://arxiv.org/abs/1201.0002} {arXiv:1201.0002 [astro-ph.IM]}
  \BibitemShut {NoStop}%
\bibitem [{\citenamefont {Abraham}\ \emph {et~al.}(2004)\citenamefont {Abraham}
  \emph {et~al.}}]{PierreAuger:2004naf}%
  \BibitemOpen
  \bibfield  {author} {\bibinfo {author} {\bibfnamefont {J.}~\bibnamefont
  {Abraham}} \emph {et~al.} (\bibinfo {collaboration} {Pierre Auger}),\
  }\bibfield  {title} {\bibinfo {title} {{Properties and performance of the
  prototype instrument for the Pierre Auger Observatory}},\ }\href
  {https://doi.org/10.1016/j.nima.2003.12.012} {\bibfield  {journal} {\bibinfo
  {journal} {Nucl. Instrum. Meth. A}\ }\textbf {\bibinfo {volume} {523}},\
  \bibinfo {pages} {50} (\bibinfo {year} {2004})}\BibitemShut {NoStop}%
\bibitem [{\citenamefont {Aab}\ \emph {et~al.}(2015)\citenamefont {Aab} \emph
  {et~al.}}]{PierreAuger:2015eyc}%
  \BibitemOpen
  \bibfield  {author} {\bibinfo {author} {\bibfnamefont {A.}~\bibnamefont
  {Aab}} \emph {et~al.} (\bibinfo {collaboration} {Pierre Auger}),\ }\bibfield
  {title} {\bibinfo {title} {{The Pierre Auger Cosmic Ray Observatory}},\
  }\href {https://doi.org/10.1016/j.nima.2015.06.058} {\bibfield  {journal}
  {\bibinfo  {journal} {Nucl. Instrum. Meth. A}\ }\textbf {\bibinfo {volume}
  {798}},\ \bibinfo {pages} {172} (\bibinfo {year} {2015})},\ \Eprint
  {https://arxiv.org/abs/1502.01323} {arXiv:1502.01323 [astro-ph.IM]}
  \BibitemShut {NoStop}%
\bibitem [{\citenamefont {Teshima}\ \emph {et~al.}(1986)\citenamefont {Teshima}
  \emph {et~al.}}]{Teshima:1986rq}%
  \BibitemOpen
  \bibfield  {author} {\bibinfo {author} {\bibfnamefont {M.}~\bibnamefont
  {Teshima}} \emph {et~al.},\ }\bibfield  {title} {\bibinfo {title}
  {{Properties of 10**9-{GeV} - 10**10-{GeV} Extensive Air Showers at Core
  Distances Between 100-m and 3000-m}},\ }\href
  {https://doi.org/10.1088/0305-4616/12/10/017} {\bibfield  {journal} {\bibinfo
   {journal} {J. Phys. G}\ }\textbf {\bibinfo {volume} {12}},\ \bibinfo {pages}
  {1097} (\bibinfo {year} {1986})}\BibitemShut {NoStop}%
\bibitem [{\citenamefont {Yoshida}\ \emph {et~al.}(1994)\citenamefont {Yoshida}
  \emph {et~al.}}]{Yoshida:1994jf}%
  \BibitemOpen
  \bibfield  {author} {\bibinfo {author} {\bibfnamefont {S.}~\bibnamefont
  {Yoshida}} \emph {et~al.},\ }\bibfield  {title} {\bibinfo {title} {{Lateral
  distribution of charged particles in giant air showers above EeV observed by
  AGASA}},\ }\href {https://doi.org/10.1088/0954-3899/20/4/011} {\bibfield
  {journal} {\bibinfo  {journal} {J. Phys. G}\ }\textbf {\bibinfo {volume}
  {20}},\ \bibinfo {pages} {651} (\bibinfo {year} {1994})}\BibitemShut
  {NoStop}%
\bibitem [{\citenamefont {Takeda}\ \emph {et~al.}(2003)\citenamefont {Takeda}
  \emph {et~al.}}]{Takeda:2002at}%
  \BibitemOpen
  \bibfield  {author} {\bibinfo {author} {\bibfnamefont {M.}~\bibnamefont
  {Takeda}} \emph {et~al.},\ }\bibfield  {title} {\bibinfo {title} {{Energy
  determination in the Akeno Giant Air Shower Array experiment}},\ }\href
  {https://doi.org/10.1016/S0927-6505(02)00243-8} {\bibfield  {journal}
  {\bibinfo  {journal} {Astropart. Phys.}\ }\textbf {\bibinfo {volume} {19}},\
  \bibinfo {pages} {447} (\bibinfo {year} {2003})},\ \Eprint
  {https://arxiv.org/abs/astro-ph/0209422} {arXiv:astro-ph/0209422}
  \BibitemShut {NoStop}%
\bibitem [{\citenamefont {Abu-Zayyad}\ \emph {et~al.}(2014)\citenamefont
  {Abu-Zayyad} \emph {et~al.}}]{TelescopeArray:2014nxa}%
  \BibitemOpen
  \bibfield  {author} {\bibinfo {author} {\bibfnamefont {T.}~\bibnamefont
  {Abu-Zayyad}} \emph {et~al.} (\bibinfo {collaboration} {Telescope Array}),\
  }\bibfield  {title} {\bibinfo {title} {{CORSIKA Simulation of the Telescope
  Array Surface Detector}},\ }\href@noop {} {\  (\bibinfo {year} {2014})},\
  \Eprint {https://arxiv.org/abs/1403.0644} {arXiv:1403.0644 [astro-ph.IM]}
  \BibitemShut {NoStop}%
\bibitem [{\citenamefont {Stokes}\ \emph {et~al.}(2012)\citenamefont {Stokes},
  \citenamefont {Cady}, \citenamefont {Ivanov}, \citenamefont {Matthews},\ and\
  \citenamefont {Thomson}}]{Stokes:2011wf}%
  \BibitemOpen
  \bibfield  {author} {\bibinfo {author} {\bibfnamefont {B.~T.}\ \bibnamefont
  {Stokes}}, \bibinfo {author} {\bibfnamefont {R.}~\bibnamefont {Cady}},
  \bibinfo {author} {\bibfnamefont {D.}~\bibnamefont {Ivanov}}, \bibinfo
  {author} {\bibfnamefont {J.~N.}\ \bibnamefont {Matthews}},\ and\ \bibinfo
  {author} {\bibfnamefont {G.~B.}\ \bibnamefont {Thomson}},\ }\bibfield
  {title} {\bibinfo {title} {{Dethinning Extensive Air Shower Simulations}},\
  }\href {https://doi.org/10.1016/j.astropartphys.2012.03.004} {\bibfield
  {journal} {\bibinfo  {journal} {Astropart. Phys.}\ }\textbf {\bibinfo
  {volume} {35}},\ \bibinfo {pages} {759} (\bibinfo {year} {2012})},\ \Eprint
  {https://arxiv.org/abs/1104.3182} {arXiv:1104.3182 [astro-ph.IM]}
  \BibitemShut {NoStop}%
\bibitem [{Note1()}]{Note1}%
  \BibitemOpen
  \bibinfo {note} {The MC described here uses events generated by the CORSIKA
  simulation package \cite {Heck:1998vt} using the QGSJET-II-03 high-energy
  hadronic interaction model \cite {Ostapchenko:2004ss} with an assumption of
  proton primaries. Since, in the end, we normalize the energy scale to that of
  the FD, the spectrum we calculate here is insensitive to the assumption of
  primary particles or the use of various available hadronic interaction
  models.}\BibitemShut {Stop}%
\bibitem [{\citenamefont {Aab}\ \emph {et~al.}(2020)\citenamefont {Aab} \emph
  {et~al.}}]{PierreAuger:2020qqz}%
  \BibitemOpen
  \bibfield  {author} {\bibinfo {author} {\bibfnamefont {A.}~\bibnamefont
  {Aab}} \emph {et~al.} (\bibinfo {collaboration} {Pierre Auger}),\ }\bibfield
  {title} {\bibinfo {title} {{Measurement of the cosmic-ray energy spectrum
  above $2.5{\times} 10^{18}$ eV using the Pierre Auger Observatory}},\ }\href
  {https://doi.org/10.1103/PhysRevD.102.062005} {\bibfield  {journal} {\bibinfo
   {journal} {Phys. Rev. D}\ }\textbf {\bibinfo {volume} {102}},\ \bibinfo
  {pages} {062005} (\bibinfo {year} {2020})},\ \Eprint
  {https://arxiv.org/abs/2008.06486} {arXiv:2008.06486 [astro-ph.HE]}
  \BibitemShut {NoStop}%
\bibitem [{\citenamefont {Kim}\ \emph {et~al.}(2023{\natexlab{a}})\citenamefont
  {Kim}, \citenamefont {Ivanov}, \citenamefont {Jui},\ and\ \citenamefont
  {Thomson}}]{Kim:2023eul}%
  \BibitemOpen
  \bibfield  {author} {\bibinfo {author} {\bibfnamefont {J.}~\bibnamefont
  {Kim}}, \bibinfo {author} {\bibfnamefont {D.}~\bibnamefont {Ivanov}},
  \bibinfo {author} {\bibfnamefont {C.}~\bibnamefont {Jui}},\ and\ \bibinfo
  {author} {\bibfnamefont {G.}~\bibnamefont {Thomson}},\ }\bibfield  {title}
  {\bibinfo {title} {{Energy Spectrum Measured by the Telescope Array Surface
  Detectors}},\ }\href {https://doi.org/10.1051/epjconf/202328302005}
  {\bibfield  {journal} {\bibinfo  {journal} {EPJ Web Conf.}\ }\textbf
  {\bibinfo {volume} {283}},\ \bibinfo {pages} {02005} (\bibinfo {year}
  {2023}{\natexlab{a}})}\BibitemShut {NoStop}%
\bibitem [{\citenamefont {Kim}\ \emph {et~al.}(2024)\citenamefont {Kim} \emph
  {et~al.}}]{TelescopeArray:2023bdy}%
  \BibitemOpen
  \bibfield  {author} {\bibinfo {author} {\bibfnamefont {J.}~\bibnamefont
  {Kim}} \emph {et~al.} (\bibinfo {collaboration} {Telescope Array}),\
  }\bibfield  {title} {\bibinfo {title} {{Highlights from the Telescope Array
  Experiment}},\ }\href {https://doi.org/10.22323/1.444.0008} {\bibfield
  {journal} {\bibinfo  {journal} {PoS}\ }\textbf {\bibinfo {volume}
  {ICRC2023}},\ \bibinfo {pages} {008} (\bibinfo {year} {2024})}\BibitemShut
  {NoStop}%
\bibitem [{\citenamefont {Verzi}\ \emph {et~al.}(2017)\citenamefont {Verzi},
  \citenamefont {Ivanov},\ and\ \citenamefont {Tsunesada}}]{Verzi:2017hro}%
  \BibitemOpen
  \bibfield  {author} {\bibinfo {author} {\bibfnamefont {V.}~\bibnamefont
  {Verzi}}, \bibinfo {author} {\bibfnamefont {D.}~\bibnamefont {Ivanov}},\ and\
  \bibinfo {author} {\bibfnamefont {Y.}~\bibnamefont {Tsunesada}},\ }\bibfield
  {title} {\bibinfo {title} {{Measurement of Energy Spectrum of Ultra-High
  Energy Cosmic Rays}},\ }\href {https://doi.org/10.1093/ptep/ptx082}
  {\bibfield  {journal} {\bibinfo  {journal} {PTEP}\ }\textbf {\bibinfo
  {volume} {2017}},\ \bibinfo {pages} {12A103} (\bibinfo {year} {2017})},\
  \Eprint {https://arxiv.org/abs/1705.09111} {arXiv:1705.09111 [astro-ph.HE]}
  \BibitemShut {NoStop}%
\bibitem [{\citenamefont {Deligny}(2020)}]{Deligny:2020gzq}%
  \BibitemOpen
  \bibfield  {author} {\bibinfo {author} {\bibfnamefont {O.}~\bibnamefont
  {Deligny}} (\bibinfo {collaboration} {Pierre Auger, Telescope Array}),\
  }\bibfield  {title} {\bibinfo {title} {{The energy spectrum of ultra-high
  energy cosmic rays measured at the Pierre Auger Observatory and at the
  Telescope Array}},\ }\href {https://doi.org/10.22323/1.358.0234} {\bibfield
  {journal} {\bibinfo  {journal} {PoS}\ }\textbf {\bibinfo {volume}
  {ICRC2019}},\ \bibinfo {pages} {234} (\bibinfo {year} {2020})},\ \Eprint
  {https://arxiv.org/abs/2001.08811} {arXiv:2001.08811 [astro-ph.HE]}
  \BibitemShut {NoStop}%
\bibitem [{\citenamefont {Abbasi}\ \emph
  {et~al.}(2021{\natexlab{a}})\citenamefont {Abbasi} \emph
  {et~al.}}]{TelescopeArray:2021zox}%
  \BibitemOpen
  \bibfield  {author} {\bibinfo {author} {\bibfnamefont {R.}~\bibnamefont
  {Abbasi}} \emph {et~al.} (\bibinfo {collaboration} {Telescope Array, Pierre
  Auger}),\ }\bibfield  {title} {\bibinfo {title} {{Joint analysis of the
  energy spectrum of ultra-high-energy cosmic rays as measured at the Pierre
  Auger Observatory and the Telescope Array}},\ }\href
  {https://doi.org/10.22323/1.395.0337} {\bibfield  {journal} {\bibinfo
  {journal} {PoS}\ }\textbf {\bibinfo {volume} {ICRC2021}},\ \bibinfo {pages}
  {337} (\bibinfo {year} {2021}{\natexlab{a}})}\BibitemShut {NoStop}%
\bibitem [{\citenamefont {Tsunesada}(2024)}]{Tsunesada:2023yhw}%
  \BibitemOpen
  \bibfield  {author} {\bibinfo {author} {\bibfnamefont {Y.}~\bibnamefont
  {Tsunesada}},\ }\bibfield  {title} {\bibinfo {title} {{Measurement of UHECR
  energy spectrum with the Pierre Auger Observatory and the Telescope Array}},\
  }\href {https://doi.org/10.22323/1.444.0406} {\bibfield  {journal} {\bibinfo
  {journal} {PoS}\ }\textbf {\bibinfo {volume} {ICRC2023}},\ \bibinfo {pages}
  {406} (\bibinfo {year} {2024})}\BibitemShut {NoStop}%
\bibitem [{\citenamefont {Patrignani}(2016)}]{Patrignani_2016}%
  \BibitemOpen
  \bibfield  {author} {\bibinfo {author} {\bibfnamefont {C.}~\bibnamefont
  {Patrignani}},\ }\bibfield  {title} {\bibinfo {title} {Review of particle
  physics},\ }\href {https://doi.org/10.1088/1674-1137/40/10/100001} {\bibfield
   {journal} {\bibinfo  {journal} {Chinese Physics C}\ }\textbf {\bibinfo
  {volume} {40}},\ \bibinfo {pages} {100001} (\bibinfo {year}
  {2016})}\BibitemShut {NoStop}%
\bibitem [{\citenamefont {Sommers}(2001)}]{Sommers:2000us}%
  \BibitemOpen
  \bibfield  {author} {\bibinfo {author} {\bibfnamefont {P.}~\bibnamefont
  {Sommers}},\ }\bibfield  {title} {\bibinfo {title} {{Cosmic ray anisotropy
  analysis with a full-sky observatory}},\ }\href
  {https://doi.org/10.1016/S0927-6505(00)00130-4} {\bibfield  {journal}
  {\bibinfo  {journal} {Astropart. Phys.}\ }\textbf {\bibinfo {volume} {14}},\
  \bibinfo {pages} {271} (\bibinfo {year} {2001})},\ \Eprint
  {https://arxiv.org/abs/astro-ph/0004016} {arXiv:astro-ph/0004016}
  \BibitemShut {NoStop}%
\bibitem [{\citenamefont {Abbasi}\ \emph {et~al.}(2014)\citenamefont {Abbasi}
  \emph {et~al.}}]{TelescopeArray:2014tsd}%
  \BibitemOpen
  \bibfield  {author} {\bibinfo {author} {\bibfnamefont {R.~U.}\ \bibnamefont
  {Abbasi}} \emph {et~al.} (\bibinfo {collaboration} {Telescope Array}),\
  }\bibfield  {title} {\bibinfo {title} {{Indications of Intermediate-Scale
  Anisotropy of Cosmic Rays with Energy Greater Than 57 EeV in the Northern Sky
  Measured with the Surface Detector of the Telescope Array Experiment}},\
  }\href {https://doi.org/10.1088/2041-8205/790/2/L21} {\bibfield  {journal}
  {\bibinfo  {journal} {Astrophys. J. Lett.}\ }\textbf {\bibinfo {volume}
  {790}},\ \bibinfo {pages} {L21} (\bibinfo {year} {2014})},\ \Eprint
  {https://arxiv.org/abs/1404.5890} {arXiv:1404.5890 [astro-ph.HE]}
  \BibitemShut {NoStop}%
\bibitem [{\citenamefont {Abbasi}\ \emph
  {et~al.}(2021{\natexlab{b}})\citenamefont {Abbasi} \emph
  {et~al.}}]{TelescopeArray:2021dfb}%
  \BibitemOpen
  \bibfield  {author} {\bibinfo {author} {\bibfnamefont {R.~U.}\ \bibnamefont
  {Abbasi}} \emph {et~al.} (\bibinfo {collaboration} {Telescope Array}),\
  }\bibfield  {title} {\bibinfo {title} {{Indications of a Cosmic Ray Source in
  the Perseus-Pisces Supercluster}},\ }\href@noop {} {\  (\bibinfo {year}
  {2021}{\natexlab{b}})},\ \Eprint {https://arxiv.org/abs/2110.14827}
  {arXiv:2110.14827 [astro-ph.HE]} \BibitemShut {NoStop}%
\bibitem [{\citenamefont {Kim}\ \emph {et~al.}(2023{\natexlab{b}})\citenamefont
  {Kim}, \citenamefont {Ivanov}, \citenamefont {Kawata}, \citenamefont
  {Sagawa},\ and\ \citenamefont {Thomson}}]{Kim:2023ksw}%
  \BibitemOpen
  \bibfield  {author} {\bibinfo {author} {\bibfnamefont {J.}~\bibnamefont
  {Kim}}, \bibinfo {author} {\bibfnamefont {D.}~\bibnamefont {Ivanov}},
  \bibinfo {author} {\bibfnamefont {K.}~\bibnamefont {Kawata}}, \bibinfo
  {author} {\bibfnamefont {H.}~\bibnamefont {Sagawa}},\ and\ \bibinfo {author}
  {\bibfnamefont {G.}~\bibnamefont {Thomson}} (\bibinfo {collaboration}
  {Telescope Array}),\ }\bibfield  {title} {\bibinfo {title} {{Anisotropies in
  the arrival direction distribution of ultra-high energy cosmic rays measured
  by the Telescope Array surface detector}},\ }\href
  {https://doi.org/10.22323/1.444.0244} {\bibfield  {journal} {\bibinfo
  {journal} {PoS}\ }\textbf {\bibinfo {volume} {ICRC2023}},\ \bibinfo {pages}
  {244} (\bibinfo {year} {2023}{\natexlab{b}})}\BibitemShut {NoStop}%
\bibitem [{Note2()}]{Note2}%
  \BibitemOpen
  \bibinfo {note} {Note that the anisotropy signal regions depicted are based
  on the previous analysis results as follows. The Hotspot was identified in
  events with energies exceeding $\sim $$10^{19.75}$~eV, located at equatorial
  coordinates ($144.0^{\circ }, 40.5^{\circ }$) within a $25^{\circ }$ radius.
  Additionally, we observed additional anisotropies in events with energies
  greater than $10^{19.4}$~eV in the direction of the Perseus-Pisces
  supercluster. The PPSC excess was located at ($17.9^{\circ }, 35.2^{\circ }$)
  in equatorial coordinates within a $20^{\circ }$ radius.}\BibitemShut {Stop}%
\bibitem [{\citenamefont {Abreu}\ \emph {et~al.}(2022)\citenamefont {Abreu}
  \emph {et~al.}}]{PierreAuger:2022axr}%
  \BibitemOpen
  \bibfield  {author} {\bibinfo {author} {\bibfnamefont {P.}~\bibnamefont
  {Abreu}} \emph {et~al.} (\bibinfo {collaboration} {Pierre Auger}),\
  }\bibfield  {title} {\bibinfo {title} {{Arrival Directions of Cosmic Rays
  above 32 EeV from Phase One of the Pierre Auger Observatory}},\ }\href
  {https://doi.org/10.3847/1538-4357/ac7d4e} {\bibfield  {journal} {\bibinfo
  {journal} {Astrophys. J.}\ }\textbf {\bibinfo {volume} {935}},\ \bibinfo
  {pages} {170} (\bibinfo {year} {2022})},\ \Eprint
  {https://arxiv.org/abs/2206.13492} {arXiv:2206.13492 [astro-ph.HE]}
  \BibitemShut {NoStop}%
\bibitem [{\citenamefont {Heck}\ \emph {et~al.}(1998)\citenamefont {Heck},
  \citenamefont {Knapp}, \citenamefont {Capdevielle}, \citenamefont {Schatz},\
  and\ \citenamefont {Thouw}}]{Heck:1998vt}%
  \BibitemOpen
  \bibfield  {author} {\bibinfo {author} {\bibfnamefont {D.}~\bibnamefont
  {Heck}}, \bibinfo {author} {\bibfnamefont {J.}~\bibnamefont {Knapp}},
  \bibinfo {author} {\bibfnamefont {J.~N.}\ \bibnamefont {Capdevielle}},
  \bibinfo {author} {\bibfnamefont {G.}~\bibnamefont {Schatz}},\ and\ \bibinfo
  {author} {\bibfnamefont {T.}~\bibnamefont {Thouw}},\ }\bibfield  {title}
  {\bibinfo {title} {{CORSIKA: A Monte Carlo code to simulate extensive air
  showers}},\ }\href@noop {} {\  (\bibinfo {year} {1998})}\BibitemShut
  {NoStop}%
\bibitem [{\citenamefont {Ostapchenko}(2006)}]{Ostapchenko:2004ss}%
  \BibitemOpen
  \bibfield  {author} {\bibinfo {author} {\bibfnamefont {S.}~\bibnamefont
  {Ostapchenko}},\ }\bibfield  {title} {\bibinfo {title} {{QGSJET-II: Towards
  reliable description of very high energy hadronic interactions}},\ }\href
  {https://doi.org/10.1016/j.nuclphysbps.2005.07.026} {\bibfield  {journal}
  {\bibinfo  {journal} {Nucl. Phys. B Proc. Suppl.}\ }\textbf {\bibinfo
  {volume} {151}},\ \bibinfo {pages} {143} (\bibinfo {year} {2006})},\ \Eprint
  {https://arxiv.org/abs/hep-ph/0412332} {arXiv:hep-ph/0412332} \BibitemShut
  {NoStop}%
\end{thebibliography}
%

\end{document}